\documentclass[aps,prb,twocolumn,longbibliography,floatfix,showpacs=false]{revtex4-1}
\usepackage{amsmath,amssymb,bm,graphicx}
\usepackage[colorlinks=true,linkcolor=blue,citecolor=blue,urlcolor=blue]{hyperref}

\newcommand{\EZ}{E_Z}
\newcommand{\kF}{k_F}
\newcommand{\kf}{k_f}
\newcommand{\vf}{v_f}
\newcommand{\ef}{\varepsilon_f}
\newcommand{\qTF}{q_{\rm TF}}

\setcitestyle{numbers,square,comma,sort&compress}

\begin{document}

\title{Tunneling signatures of the Lifshitz transition in a superconducting vortex}

\author{Alex Levchenko}
\affiliation{Department of Physics, University of Wisconsin--Madison, Madison, Wisconsin 53706, USA}

\date{November 23, 2026}

\begin{abstract}
The Caroli--de Gennes--Matricon states bound to a vortex line in a clean type-II superconductor form one-dimensional minibands that disperse along the vortex. The lowest miniband is separated from the Fermi level by a minigap $E_0\sim\Delta^2/E_F$. A magnetic field shifts the minibands of the two spin projections in opposite directions. At the threshold field $B^*$ the Zeeman energy equals $E_0$. Estimates for FeSe place $B^*$ at a few tesla.
The lowest miniband of one spin projection then crosses the Fermi level, and a spin-polarized one-dimensional Fermi sea forms inside the vortex core. We develop the theory of tunneling spectra across this Lifshitz transition. We solve the Bogoliubov--de Gennes equations numerically in the quantum limit for all momenta along the vortex and compute the local density of states with thermal and instrumental broadening. The zero-bias conductance at the vortex center is absent below $B^*$. It has a maximum near $B^*$ with the inverse-square-root field dependence of a one-dimensional van Hove singularity and decreases at higher fields. The next miniband produces a second onset at a finite distance from the center. The zero-bias signal is carried by one spin projection at the center and by the opposite one in the surrounding ring. We then project the screened electron interaction onto the filled miniband. The resulting Luttinger parameter is close to unity, so the core hosts a weakly interacting one-dimensional Fermi gas. Its quasiparticles relax through three-particle collisions in analogy with thermalization in quantum wires and at quantum Hall edges. The corresponding relaxation rates are established analytically. 
\end{abstract}

\maketitle

\section{Introduction}
\label{sec:intro}

Caroli, de Gennes, and Matricon (CdGM) showed that a vortex line in a clean type-II superconductor binds quasiparticle states with energies below the bulk gap $\Delta$ \cite{CdGM64}. The states are labeled by a half-odd-integer angular momentum $\mu$ and by the momentum $k_z$ along the vortex. Their energies are $\varepsilon_\mu=\mu\omega_0$, with a level spacing $\omega_0\sim\Delta^2/E_F$ that is small compared with $\Delta$ in conventional superconductors with a large Fermi energy $E_F$ \cite{CdGM64,Bardeen69,KramerPesch74}. The lowest level lies at $\omega_0/2$, so the spectrum of an isolated vortex has a minigap. When temperature or disorder broadening exceeds $\omega_0$, the levels merge into a continuum, and the core responds like a normal metal \cite{BardeenStephen65}.

Hess and co-workers gave the first direct evidence for the core states. Their scanning tunneling microscopy (STM) experiments on NbSe$_2$ found a zero-bias conductance peak at the vortex center \cite{Hess89}. The peak splits with distance from the center, and the conductance maps have a sixfold star shape \cite{Hess90}. The theory explained the zero-bias peak as the envelope of many unresolved CdGM levels \cite{Shore89,Gygi91}, and vortex-core spectroscopy became a standard application of STM \cite{Fischer07,Suderow14}. Hayashi \emph{et al.} considered the quantum limit, in which the temperature is smaller than $\omega_0$ \cite{Hayashi98}. They predicted discrete peaks at finite bias, a particle-hole asymmetric spectrum at the vortex center, and Friedel-like oscillations of the local density of states (LDOS). Observation of these features requires a large ratio $\Delta/E_F$. Signatures of the quantum limit were reported for YNi$_2$B$_2$C \cite{Kaneko12}. Discrete CdGM levels were then resolved in iron-based superconductors, including FeTe$_{0.55}$Se$_{0.45}$ \cite{Chen18}, (Li,Fe)OHFeSe \cite{Liu18}, monolayer FeSe on SrTiO$_3$ \cite{ChenC20}, and bulk FeSe \cite{Hanaguri19}. The search for Majorana zero modes in the vortices of Fe(Te,Se) \cite{Wang18,Machida19,Kong19,Zhu20,MachidaHanaguri24} stimulated the development of dilution-refrigerator STM with an energy resolution of about $20\,\mu$eV in fields up to $17.5$\,T \cite{Machida18}.

In a three-dimensional superconductor each CdGM level is a one-dimensional (1D) miniband $\varepsilon_\mu(k_z)$. Makhlin and Volovik pointed out that the Zeeman energy shifts the minibands of the two spin projections in opposite directions \cite{MakhlinVolovik95}. When the Zeeman energy exceeds the minigap, the lowest miniband of one spin projection crosses the Fermi level at two points $\pm\kf$. Fermions with negative energy then form a 1D Fermi liquid inside the core. They further studied the instability of this liquid toward a spin-triplet component of the order parameter in the core and estimated its transition temperature \cite{MakhlinVolovik95}. Vishwanath and Senthil later described the filled miniband as a spin-polarized Luttinger liquid and proposed a power-law tunneling density of states as its signature \cite{VishwanathSenthil01}. The appearance of Fermi points in a band at a critical value of a control parameter is a Lifshitz transition \cite{Lifshitz60,Volovik17}. Here the transition takes place inside the vortex core, and the control parameter is the magnetic field. Closely related physics is known for vortices in spin-imbalanced atomic Fermi gases, where the core states are occupied by the majority species \cite{Takahashi06,HuLiuDrummond07}, and for superconductors with strong Pauli paramagnetism, where the quasiclassical theory gives a paramagnetic moment in the core \cite{IchiokaMachida07}.

Several questions remained open in these works. First, the tunneling spectrum across the transition was not calculated. The magnitude and the field dependence of the zero-bias signal, its spatial and spin structure, and the energy resolution needed to detect it are not known. Second, the strength of the interaction between the quasiparticles of the filled miniband was not evaluated, so it is not known whether the Luttinger-liquid exponents differ noticeably from those of free fermions. Third, the relaxation of quasiparticles in the filled miniband was not considered. Fourth, no material was identified in which the threshold field lies below the upper critical field and the minigap exceeds the experimental resolution.

In this paper we address these questions. We solve the Bogoliubov--de Gennes (BdG) equations numerically in the quantum limit for all $k_z$ and for both spin projections, and we compute the tunneling conductance as a function of bias, position, and field. The calculation shows that the transition has a distinct signature in STM. The lowest peak at the vortex center splits linearly with the field. One of its branches reaches zero bias at the threshold field $B^*$. The zero-bias conductance at the center vanishes below $B^*$, passes through a maximum near $B^*$, and decreases at higher fields as the inverse square root of the distance to the threshold. This field dependence reflects the van Hove singularity at the bottom of a 1D band. We also find that the zero-bias conductance is carried by spin-up electrons at the center and by spin-down electrons in the surrounding ring, and that the next miniband gives a second onset in the ring at a higher field.

We then use the BdG wave functions to project the screened interaction onto the filled miniband. The dimensionless coupling is of the order of $0.1$ or smaller for realistic screening, and the Luttinger parameter is close to unity. The filled miniband is therefore a weakly interacting 1D Fermi gas. Its kinetics is controlled by three-particle collisions, as in quantum wires \cite{Lunde07,Khodas07,Karzig10,Micklitz10,Levchenko10,MicklitzLevchenko11,Imambekov12,MatveevFurusaki13,LevchenkoMicklitz21} and at quantum Hall edges \cite{Karzig12}. We derive the three-particle amplitude for the vortex and obtain a decay rate of a hot quasiparticle that scales as the eighth power of its energy.

The transition requires that the Zeeman energy at fields below the upper critical field $H_{c2}$ exceeds the minigap. This condition favors superconductors with a large ratio $\Delta/E_F$ and a heavy band mass. FeSe satisfies it. The compound is in the crossover between the Bardeen--Cooper--Schrieffer and Bose--Einstein regimes, with $\Delta/E_F$ of the order of $0.1$ to $1$ depending on the band \cite{Kasahara14,Kasahara16,Shibauchi20}. STM at $0.25$\,T resolves the lowest core level at $E_0\approx0.2$\,meV and coherence peaks at $2.5$\,meV \cite{Hanaguri19}. For a $g$ factor of 2 this gives $B^*\approx3.5$\,T, well below $H_{c2}\approx16$\,T for a field along the $c$ axis \cite{Kasahara14}. FeSe is stoichiometric, and its band structure has no topological surface state \cite{MachidaHanaguri24}, so a field-induced zero-bias peak in its vortices cannot be attributed to a Majorana mode. Vortex-core states have also been observed in superconducting FeTe films \cite{Yan26}, and terahertz spectroscopy of FeTe$_{1-x}$Se$_x$ films resolves transitions between CdGM states and their lifetimes \cite{Room26}.

The paper is organized as follows. Section~\ref{sec:model} introduces the model, the numerical solution of the BdG equations, and the effect of the Zeeman field. Section~\ref{sec:stm} presents the tunneling spectra. Section~\ref{sec:int} evaluates the interaction in the filled miniband. Section~\ref{sec:kin} treats relaxation by three-particle collisions and its relation to earlier work on vortex dynamics. Section~\ref{sec:sum} summarizes the results. Appendix~\ref{app:num} gives numerical details, and Appendix~\ref{app:three} describes the calculation of the three-particle amplitude and of the rates.

\begin{figure}[t!]
\includegraphics[width=0.92\columnwidth]{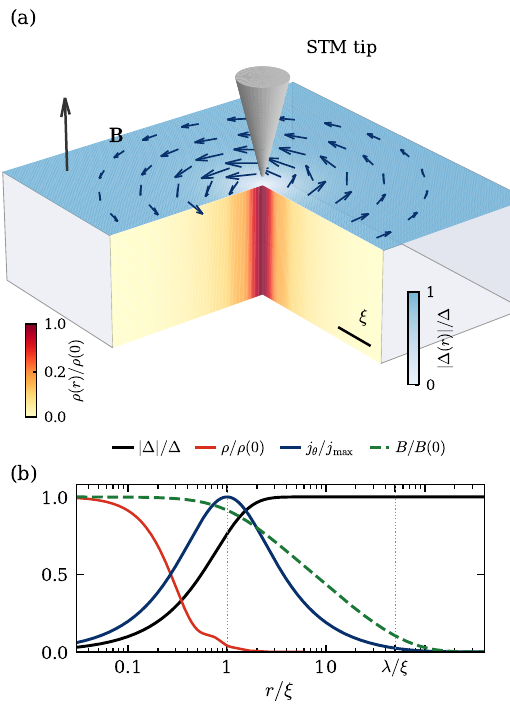}
\caption{Geometry and length scales. (a) Cutaway view of a vortex line along the magnetic field $\mathbf{B}$, drawn to scale in units of $\xi$. The color of the top surface shows the gap profile $|\Delta(r)|=\Delta\tanh(r/\xi)$ used in this work. The color of the cut faces shows the probability density $\rho=f^2+g^2$ of the lowest bound state, obtained from the BdG equations for $k_F\xi=6$ at $k_z=0$. The arrows show the supercurrent, and their length is proportional to the current density $j_\theta(r)$. An STM tip above the core measures the local density of states at the surface. (b) Radial profiles of the gap, of the density $\rho$, and of the supercurrent density and the magnetic field in the Clem model \cite{Clem75} for $\lambda/\xi=50$, where $\lambda$ is the penetration depth.}
\label{fig:schematic}
\end{figure}

\section{Model and vortex-core minibands}
\label{sec:model}

\subsection{Bogoliubov--de Gennes equations}
We consider a clean $s$-wave superconductor with a parabolic band of mass $m$, Fermi momentum $k_F$, and Fermi energy $E_F=\hbar^2k_F^2/2m$. A single vortex is directed along the $z$ axis, which is also the direction of the field (Fig.~\ref{fig:schematic}). Figure~\ref{fig:schematic}(b) shows the length scales of the problem. The gap recovers at $r\sim\xi$. The supercurrent density has a maximum at $r\approx\xi$ and decays as $1/r$ up to the penetration depth $\lambda$. The lowest bound state is confined to distances of the order of $1/k_F$ from the axis. We consider a strongly type-II superconductor, $\lambda\gg\xi$, and neglect the vector potential of the vortex in the equations for the core states. The order parameter is $\Delta(\mathbf{r})=\Delta(r)e^{-i\theta}$ in cylindrical coordinates $(r,\theta,z)$. In the Nambu basis $(\psi_\uparrow,\psi_\downarrow^\dagger)$ the BdG equations read
\begin{equation}
\begin{pmatrix}\hat\xi & \Delta(\mathbf{r})\\ \Delta^*(\mathbf{r}) & -\hat\xi\end{pmatrix}
\begin{pmatrix}u\\ v\end{pmatrix}=E\begin{pmatrix}u\\ v\end{pmatrix},\qquad
\hat\xi=-\frac{\hbar^2\nabla^2}{2m}-E_F .
\label{eq:bdg}
\end{equation}
The solutions have the form
\begin{equation}
\begin{pmatrix}u\\ v\end{pmatrix}=\frac{e^{ik_zz}}{\sqrt{L}}
\begin{pmatrix}e^{i(\mu-\frac12)\theta}f(r)\\ e^{i(\mu+\frac12)\theta}g(r)\end{pmatrix},
\label{eq:ansatz}
\end{equation}
where $L$ is the length of the vortex and $\mu$ is a half-odd integer. The radial functions satisfy
\begin{subequations}\label{eq:radial}
\begin{align}
-\frac{\hbar^2}{2m}\Big[f''+\frac{f'}{r}-\frac{(\mu-\frac12)^2}{r^2}f\Big]-E_\perp f+\Delta(r)g&=Ef,\\
\frac{\hbar^2}{2m}\Big[g''+\frac{g'}{r}-\frac{(\mu+\frac12)^2}{r^2}g\Big]+E_\perp g+\Delta(r)f&=Eg,
\end{align}
\end{subequations}
with the normalization $2\pi\int_0^\infty r\,dr\,(f^2+g^2)=1$. The motion along the vortex enters only through the transverse Fermi energy
\begin{equation}
E_\perp(k_z)=E_F\left(1-\frac{k_z^2}{\kF^2}\right).
\label{eq:Eperp}
\end{equation}
Each value of $k_z$ therefore defines a two-dimensional vortex problem with the transverse Fermi momentum $k_\perp=(\kF^2-k_z^2)^{1/2}$. Equations~\eqref{eq:radial} are invariant under $(E,\mu,f,g)\to(-E,-\mu,g,-f)$, so the eigenstates come in pairs with opposite energies.

We use the profile $\Delta(r)=\Delta\tanh(r/\xi)$ with $\xi=\hbar v_F/\Delta$ and $v_F=\hbar\kF/m$. When energies are measured in units of $\Delta$ and lengths in units of $\xi$, Eqs.~\eqref{eq:radial} contain the single parameter $\kF\xi=2E_F/\Delta$ and depend on the momentum through the ratio $k_z/\kF$. All results below are presented in dimensionless form. The natural energy scale for the low-energy spectra is the minigap
\begin{equation}
E_0\equiv\varepsilon_{1/2}(k_z=0),
\end{equation}
and we express bias voltages, broadening, and Zeeman energies in units of $E_0$.

\begin{figure*}[t!]
\includegraphics[width=0.98\textwidth]{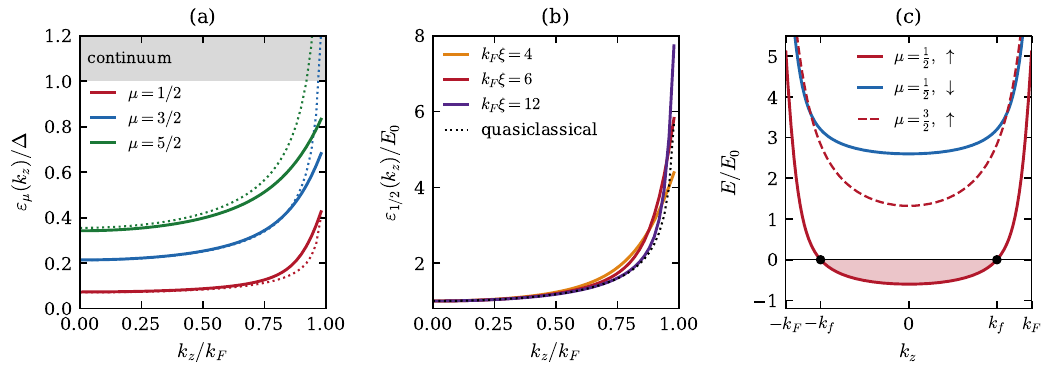}
\caption{Vortex-core minibands from the numerical solution of the BdG equations. (a) The three lowest minibands for $\kF\xi=6$ in units of the bulk gap $\Delta$. Dotted lines show the quasiclassical result \eqref{eq:qc} for the same gap profile. (b) The lowest miniband in units of the minigap $E_0$ for three values of $\kF\xi$. The dotted line is the quasiclassical result. (c) Energies of spin-up and spin-down quasiparticles for $\kF\xi=6$ at $B=1.6\,B^*$. The spin-up branch of the lowest miniband is occupied between the Fermi points $\pm\kf$ (shaded).}
\label{fig:bands}
\end{figure*}

\subsection{Minibands}

In the quasiclassical limit $\kF\xi\gg1$ the bound-state energies at $|E|\ll\Delta$ are $\varepsilon_\mu(k_z)=\mu\,\omega_0(k_z)$ with \cite{CdGM64,KramerPesch74}
\begin{equation}
\omega_0(k_z)=\frac{1}{k_\perp}\,
\frac{\int_0^\infty dr\,r^{-1}\Delta(r)\,e^{-2K(r)}}{\int_0^\infty dr\,e^{-2K(r)}},
\label{eq:qc}
\end{equation}
where $K(r)=(m/\hbar^2k_\perp)\int_0^r\Delta(r')\,dr'$.
For the adopted profile this gives $\omega_0(k_z)=(\Delta/\kF\xi)\,\gamma\,I(\gamma)$, where $\gamma=\kF/k_\perp$ and $I(\gamma)$ is the average of $\tanh(s)/s$ with the weight $\cosh^{-2\gamma}s$. At $k_z=0$ one finds $I(1)=0.853$ and $\omega_0=0.43\,\Delta^2/E_F$. The dependence on $k_z$ has two sources. The prefactor $1/k_\perp$ increases with $|k_z|$. The bound state also becomes more compact, because its radial extent is $\hbar^2k_\perp/m\Delta$, and it then probes a region of smaller $\Delta(r)$. For a profile of fixed width the second effect partially compensates the first one, and $\omega_0\propto(1-k_z^2/\kF^2)^{-1/2}$ as $|k_z|\to\kF$. The frequently quoted form $\omega_0\propto\Delta^2/E_\perp\propto(1-k_z^2/\kF^2)^{-1}$ applies when the core radius scales with the extent of the bound state.

The quantum limit corresponds to moderate values of $\kF\xi$, for which the quasiclassical approximation is not controlled. We therefore solve Eqs.~\eqref{eq:radial} numerically (Appendix~\ref{app:num}). Figure~\ref{fig:bands}(a) shows the three lowest minibands for $\kF\xi=6$. The energies at $k_z=0$ are $E_0=0.073\,\Delta$, $\varepsilon_{3/2}(0)=2.92\,E_0$, and $\varepsilon_{5/2}(0)=4.68\,E_0$. The ladder is not equidistant because the higher levels approach the gap edge. The quasiclassical expression \eqref{eq:qc} with the same profile, shown by dotted lines, reproduces $E_0$ within $3\%$ and overestimates the dispersion at large $|k_z|$. The lowest miniband has the expansion
\begin{equation}
\frac{\varepsilon_{1/2}(k_z)}{E_0}=1+0.61\,\frac{k_z^2}{\kF^2}+0.52\,\frac{k_z^4}{\kF^4}+\dots
\label{eq:band}
\end{equation}
and reaches $6.7\,E_0\approx0.49\,\Delta$ at $|k_z|\to\kF$. Its mass at the bottom is
\begin{equation}
m^*=\frac{\hbar^2\kF^2}{1.22\,E_0}=1.63\,\frac{E_F}{E_0}\,m\approx 67\,m .
\label{eq:mstar}
\end{equation}
Figure~\ref{fig:bands}(b) compares the lowest miniband for $\kF\xi=4$, $6$, and $12$ in units of $E_0$. The curves are close to each other for $|k_z|\lesssim0.7\,\kF$, which shows that $E_0$ and $\kF$ are the appropriate scales.

\begin{figure*}[t!]
\includegraphics[width=0.98\textwidth]{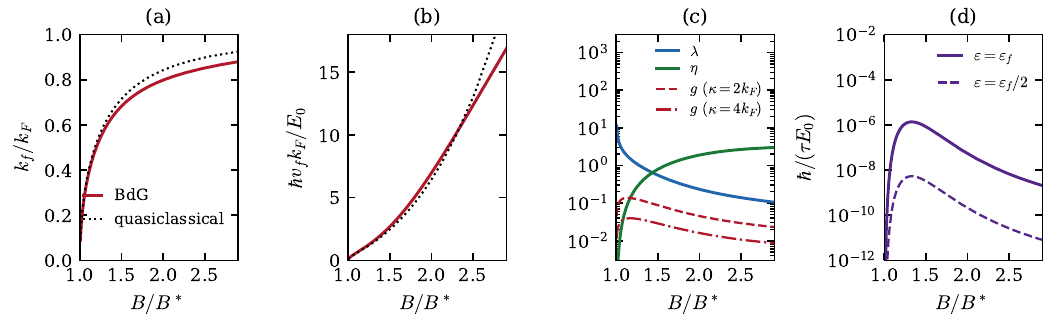}
\caption{Parameters of the filled miniband as functions of field for $\kF\xi=6$. (a) Fermi momentum $\kf$ from the numerical dispersion and from the quasiclassical formula \eqref{eq:qc}. (b) Fermi velocity. (c) Interaction parameters $\lambda$ [Eq.~\eqref{eq:lambda}], $\eta$ [Eq.~\eqref{eq:eta}], and $g_{1D}$ [Eq.~\eqref{eq:K}] for two values of the screening wave vector. (d) Estimate of the three-particle decay rate \eqref{eq:rateMF} of a quasiparticle with excitation energy $\varepsilon$.}
\label{fig:params}
\end{figure*}

\subsection{Zeeman field and the threshold}

The Zeeman term is $H_Z=-\EZ\int d^3r\,(\psi_\uparrow^\dagger\psi_\uparrow-\psi_\downarrow^\dagger\psi_\downarrow)$ with $\EZ=g\mu_BB/2$, where $\mu_B$ is the Bohr magneton and spin up denotes the projection with the lower Zeeman energy. In the Nambu basis of Eq.~\eqref{eq:bdg} this term equals $-\EZ$ times the unit matrix, up to a constant. For singlet pairing without spin-orbit coupling it commutes with the BdG Hamiltonian. All eigenvalues shift by $-\EZ$, and the wave functions do not change \cite{MakhlinVolovik95,VishwanathSenthil01}. A spin-up quasiparticle in the state $(\mu,k_z)$ has the energy $\varepsilon_\mu(k_z)-\EZ$, and a spin-down quasiparticle has the energy $\varepsilon_\mu(k_z)+\EZ$.

The orbital effect of the field on the core states is also linear in $B$. In the symmetric gauge the term linear in the vector potential equals $(|e|B/2m)\hat L_z$ in both the electron and the hole blocks \cite{VishwanathSenthil01}. To first order it shifts the energy of the state \eqref{eq:ansatz} by $\mu\,(m_e/m)\,\mu_BB$, where $m_e$ is the free-electron mass. We used the fact that the bound states are almost neutral, $2\pi\!\int r\,dr\,(f^2-g^2)\ll1$. For the lowest miniband the two contributions combine into
\begin{equation}
\varepsilon_{1/2}(k_z)-\tfrac12g_{\rm eff}\,\mu_BB,\qquad g_{\rm eff}=g-\frac{m_e}{m}.
\label{eq:geff}
\end{equation}
The lowest spin-up miniband reaches the Fermi level at the threshold field
\begin{equation}
B^*=\frac{2E_0}{g_{\rm eff}\,\mu_B}.
\label{eq:Bstar}
\end{equation}
For $B>B^*$ the miniband is occupied between the Fermi points $\pm\kf$ defined by
\begin{equation}
\varepsilon_{1/2}(\kf)=\EZ ,
\label{eq:kf}
\end{equation}
where $\EZ$ now includes the orbital correction of Eq.~\eqref{eq:geff}. This configuration is shown in Fig.~\ref{fig:bands}(c). The Fermi energy of the 1D gas, counted from the bottom of the miniband, and its Fermi velocity are
\begin{equation}
\ef=\EZ-E_0=E_0\left(\frac{B}{B^*}-1\right),\qquad
\vf=\frac{1}{\hbar}\frac{\partial\varepsilon_{1/2}}{\partial k_z}\Big|_{\kf}.
\label{eq:ef}
\end{equation}
Near the threshold the band is parabolic, $\kf=(2m^*\ef)^{1/2}/\hbar$ and $\vf=\hbar\kf/m^*$. The Fermi momentum and velocity obtained from the numerical dispersion are shown in Figs.~\ref{fig:params}(a) and \ref{fig:params}(b). The miniband with $\mu=3/2$ becomes occupied at a second threshold $B_2^*$. If the orbital correction is neglected, $B_2^*/B^*=\varepsilon_{3/2}(0)/E_0=2.92$. The orbital shift is three times larger for $\mu=3/2$, which increases this ratio to $2.92\,(g-m_e/m)/(g-3m_e/m)$. In the figures we use the variable $B/B^*=\EZ/E_0$ and apply the same shift to all minibands.

The model neglects the suppression of $\Delta(r)$ by the spin polarization of the core and by the field. Both effects are small when $\ef\ll\Delta$ and $B\ll H_{c2}$. We also consider an isolated vortex. The bound states decay as $e^{-r/\xi}$ outside the core. Their overlap between neighboring vortices at a distance $a_v$ is proportional to $e^{-a_v/\xi}$ and gives the levels a finite width in the vortex lattice. We neglect this width, which requires $a_v$ to be several times $\xi$.

\begin{figure*}[t!]
\includegraphics[width=0.98\textwidth]{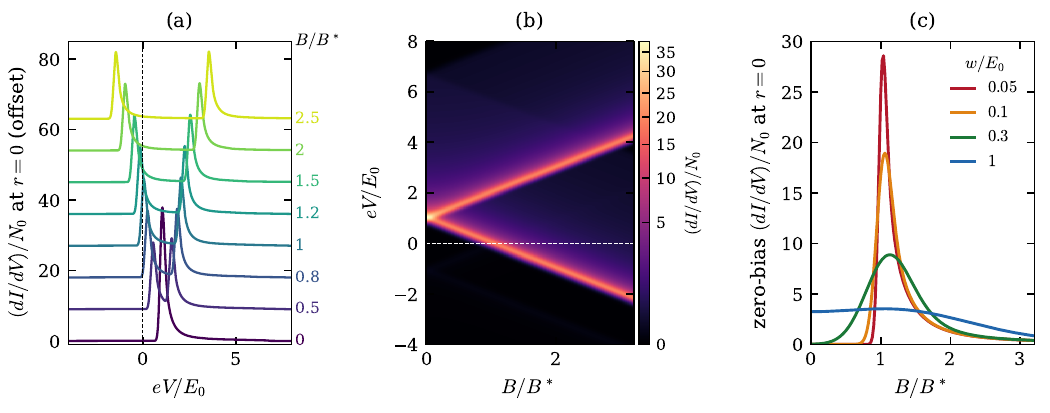}
\caption{Tunneling conductance at the vortex center for $\kF\xi=6$, normalized to the normal-state value $N_0$. (a) Spectra at several fields for the broadening $w=0.1\,E_0$. The curves are offset by $9$ units. (b) The same quantity as a function of field and bias. (c) Zero-bias conductance as a function of field for four values of the broadening.}
\label{fig:spectra}
\end{figure*}

\section{Tunneling spectra}
\label{sec:stm}

\subsection{Local density of states}

For a nonmagnetic tip and a weak tunnel coupling the differential conductance is proportional to the LDOS of electrons, thermally broadened. With the Zeeman shift included, the LDOS of electrons with spin projection $\sigma=\pm1$ is
\begin{equation}
N_\sigma(r,E)=\sum_{n}\int\frac{dk_z}{2\pi}\,f_n^2(r)\,\delta\big(E-E_n+\sigma\EZ\big).
\label{eq:ldos}
\end{equation}
The sum runs over all eigenstates $n$ of Eqs.~\eqref{eq:radial} at a given $k_z$ with both signs of the energy $E_n$ and all $\mu$. The spin-down contribution was transformed to this form with the help of the symmetry stated below Eq.~\eqref{eq:Eperp}. We normalize the total LDOS, $N=N_\uparrow+N_\downarrow$, to its normal-state value $N_0=m\kF/\pi^2\hbar^2$. Thermal smearing and the instrumental resolution are modeled by a convolution with a Gaussian of standard deviation $w$. A Gaussian with $w=1.5\,k_BT$ has the same full width at half maximum as the thermal kernel $-\partial f/\partial E$, and we use $w^2=w_{\rm inst}^2+(1.5\,k_BT)^2$ as an estimate of the combined width. The broadened quantity is denoted $dI/dV$.

Only the components with zero orbital momentum are finite at $r=0$. According to Eq.~\eqref{eq:ansatz} this selects $f$ for $\mu=1/2$. The spectrum at the vortex center therefore contains the lowest miniband alone, at positive energies. At $B=0$ this gives one peak at $eV=E_0$ with a tail toward higher bias that comes from the dispersion along $k_z$. The asymmetry between positive and negative bias is the property of the quantum limit established in Ref.~\cite{Hayashi98}. For holelike carriers the peak appears at negative bias.

\subsection{Field dependence at the vortex center}

Figure~\ref{fig:spectra}(a) shows the spectra at the vortex center for $w=0.1\,E_0$. In a field the peak splits into a spin-up component at $E_0-\EZ$ and a spin-down component at $E_0+\EZ$. The distance between them is $2\EZ=g_{\rm eff}\mu_BB$. The spin-up component reaches zero bias at $B=B^*$ and moves to negative bias at higher fields. Figure~\ref{fig:spectra}(b) presents the same data as a function of field and bias. The two straight ridges are the bottoms of the spin-up and spin-down minibands. The slope of the ridges gives $g_{\rm eff}$, and the intercept of the lower ridge with zero bias gives $B^*$. Both can be measured at fields below the threshold.

Each ridge has an asymmetric profile with an abrupt onset on the low-bias side and a tail on the high-bias side. This is the density of states of a 1D band. From Eq.~\eqref{eq:ldos}, the contribution of the lowest miniband is
\begin{equation}
N_\sigma(r,E)=\frac{f_{k}^2(r)}{\pi\hbar\,|v(k)|}\,\Big|_{\varepsilon_{1/2}(k)=E+\sigma\EZ},
\label{eq:vhs}
\end{equation}
where $f_k$ is the radial function of the state with $k_z=k$ and $v(k)=\hbar^{-1}\partial\varepsilon_{1/2}/\partial k$. Near the bottom of the band $v\propto(E+\sigma\EZ-E_0)^{1/2}$, so the LDOS has an inverse-square-root van Hove singularity. In the units of Fig.~\ref{fig:spectra},
\begin{equation}
\frac{N_\uparrow(0,E)}{N_0}=\frac{A}{\sqrt{(E+\EZ-E_0)/E_0}},\qquad E+\EZ>E_0 ,
\label{eq:vhs2}
\end{equation}
and $A=6.8$ for $\kF\xi=6$. The coefficient follows from $f_0^2(0)=1.50/\xi^2$ and from the mass \eqref{eq:mstar}.

The zero-bias conductance at the center is shown in Fig.~\ref{fig:spectra}(c). 
The decrease follows Eq.~\eqref{eq:vhs} at $E=0$,
\begin{equation}
N(r,0)=\frac{f_{\kf}^2(r)+g_{\kf}^2(r)}{\pi\hbar\vf},
\label{eq:zb}
\end{equation}
with the Fermi velocity that grows with the field. The first term in the numerator is the spin-up part and the second term is the spin-down part, as discussed below. The height of the maximum is set by the broadening. 
For $w=0.3\,E_0$ the maximum is $9\,N_0$. For $w=E_0$ the peak at $E_0$ is not resolved from zero bias already at $B=0$, and the zero-bias conductance has no feature at the threshold. The transition is therefore visible when the total broadening is a fraction of the minigap. The zero-bias conductance is not a monotonic function of the field. It has the form of a maximum with a slowly decaying tail, in contrast with a zero mode pinned at the Fermi level.

\subsection{Spatial and spin structure}

\begin{figure*}[t]
\includegraphics[width=0.98\textwidth]{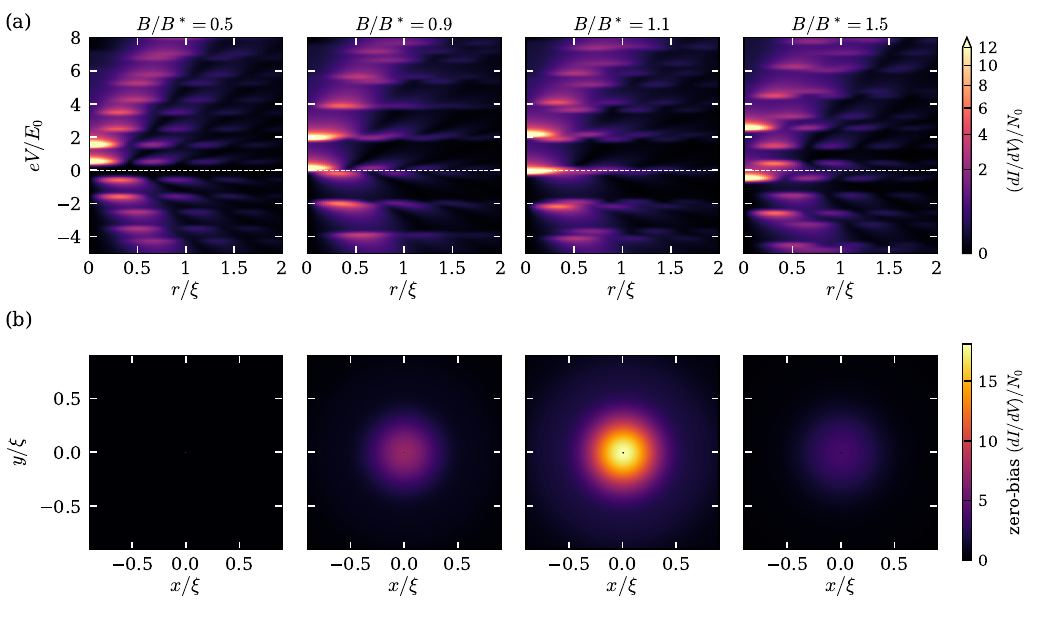}
\caption{Spatial structure of the tunneling conductance for $\kF\xi=6$ and $w=0.1\,E_0$. (a) Conductance as a function of bias and distance $r$ from the vortex center at four fields. (b) Zero-bias conductance maps at the same fields on a common color scale.}
\label{fig:maps}
\end{figure*}

\begin{figure*}[t]
\includegraphics[width=0.98\textwidth]{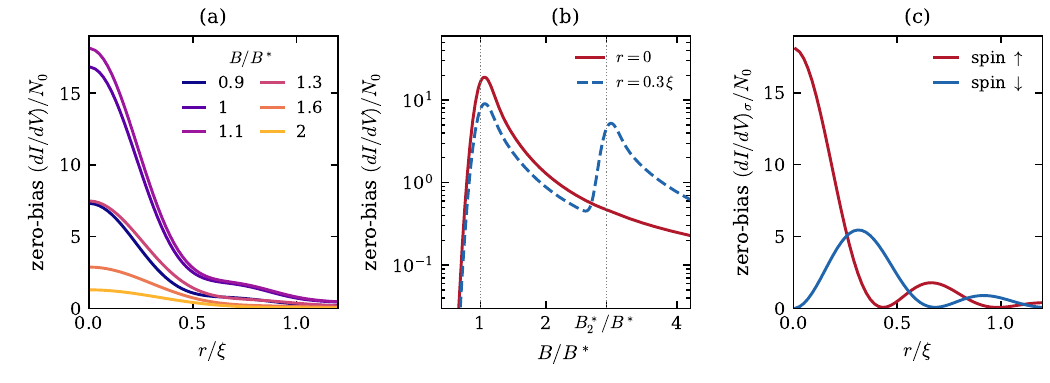}
\caption{Zero-bias conductance for $\kF\xi=6$ and $w=0.1\,E_0$. (a) Radial profiles at several fields. (b) Field dependence at the vortex center and at $r=0.3\,\xi$. The dotted vertical lines mark the thresholds $B^*$ and $B_2^*$ of the minibands with $\mu=1/2$ and $\mu=3/2$. (c) Spin-resolved radial profiles at $B=1.1\,B^*$.}
\label{fig:onsets}
\end{figure*}

Figure~\ref{fig:maps}(a) shows the conductance as a function of bias and distance from the center at four fields. At low field the spectrum consists of the levels with $\mu=1/2,3/2,\dots$ at positive bias and their particle-hole partners at negative bias. A level with a given $\mu$ has its maximum at a distance of the order of $\mu/\kF$, and the bias of the maximum increases with $r$ in steps, as observed in FeSe \cite{Hanaguri19}. The field splits every level. Near the threshold the lowest spin-up level merges with its particle-hole partner at zero bias. Figure~\ref{fig:maps}(b) shows the corresponding zero-bias maps. The map is dark at $0.5\,B^*$. A spot with a radius at half maximum of $0.29\,\xi\approx1.7/\kF$ appears as the field approaches $B^*$ and becomes weaker above the threshold. The radial profiles are given in Fig.~\ref{fig:onsets}(a).

The spin content of the zero-bias signal follows from Eq.~\eqref{eq:ldos}. A spin-up electron tunnels into the state at the Fermi point with the weight $f_{\kf}^2(r)$. A spin-down electron tunnels with the weight $g_{\kf}^2(r)$, because adding a spin-down electron removes a spin-up quasiparticle from the occupied miniband. The function $f$ has zero orbital momentum and is maximal at the center. The function $g$ has unit orbital momentum, vanishes at the center, and has its maximum at $r\approx0.3\,\xi$. The spin-resolved profiles are shown in Fig.~\ref{fig:onsets}(c). The zero-bias conductance is fully spin-up polarized at the center. The spin-down part exceeds the spin-up part between $r=0.27\,\xi$ and $0.57\,\xi$, and at larger distances the two parts alternate. The integrals of the two parts over the plane are equal within $2\%$, which is another expression of the neutrality of the bound state. A spin-polarized tip \cite{Wiesendanger09} would thus detect a reversal of the polarization of the zero-bias signal between the center and the ring. At finite bias the peaks at $E_0-\EZ$ and $E_0+\EZ$ at the center have opposite polarizations. The vortex core in a field is therefore spin polarized without any topological origin, which is relevant for proposals to identify Majorana modes by their spin polarization \cite{MachidaHanaguri24}.

The second threshold is illustrated in Fig.~\ref{fig:onsets}(b). The radial function $f$ of the miniband with $\mu=3/2$ has unit orbital momentum and vanishes at the center. The zero-bias conductance at $r=0$ therefore continues to decrease across $B_2^*$. At $r=0.3\,\xi$ it has a second maximum, where it increases from $0.45\,N_0$ to $5.2\,N_0$. The sequence of maps with increasing field is a dark core, a central spot near $B^*$, a fading spot, and a ring near $B_2^*$.

\subsection{Conditions for observation and estimates for FeSe}

The results of this section depend on the material through $E_0$, $g_{\rm eff}$, and $\kF\xi$. In FeSe the lowest level at the vortex center is observed at $E_0\approx0.2$\,meV with a larger weight at positive bias, and the coherence peaks are at $\Delta\approx2.5$\,meV \cite{Hanaguri19}. The ratio $E_0/\Delta\approx0.08$ corresponds to $\kF\xi\approx6$ in our model, which motivated this choice of the parameter. With $g_{\rm eff}=2$ Eq.~\eqref{eq:Bstar} gives $B^*=3.5$\,T, and the two components of the central peak separate at the rate $g_{\rm eff}\mu_B=0.12$\,meV/T. The band masses in FeSe are several free-electron masses \cite{Terashima14}, so the orbital correction in Eq.~\eqref{eq:geff} reduces $g_{\rm eff}$ by $0.2$ to $0.3$ and increases $B^*$ accordingly. Without the orbital correction the second threshold is at $2.92\,B^*\approx10$\,T. The orbital correction shifts it to higher fields, toward $H_{c2}$.

The broadening $w=0.1\,E_0=20\,\mu$eV corresponds to the resolution of dilution-refrigerator STM at an electron temperature below $100$\,mK \cite{Machida18,Machida19}. The value $w=0.3\,E_0$ corresponds to measurements at $0.4$\,K, for which $1.5\,k_BT=52\,\mu$eV. In that case the zero-bias conductance still has a maximum near $1.1\,B^*$, with a height of $9\,N_0$ and a finite value below the threshold. A resolution of $0.25$\,meV, typical of earlier measurements on Fe(Te,Se) \cite{Chen18,Wang18,Kong19}, corresponds to $w\approx E_0$ and does not resolve the transition.

Several effects are not included in the model. The first is the elastic width of the levels. The transition is resolved if this width is small compared with $E_0$. Terahertz measurements on FeTe$_{1-x}$Se$_x$ films give lifetimes of the CdGM states between $0.4$ and $0.8$\,ps, comparable to the inverse level spacing in those films \cite{Room26}. In FeSe single crystals the lowest level is resolved by STM \cite{Hanaguri19}, which indicates a smaller width. The second effect is the band structure. FeSe has electron and hole pockets, an anisotropic gap, and a nematic distortion that elongates the vortices \cite{Hanaguri19,Shibauchi20}. Each pocket has its own minigap and its own threshold field. For a weakly warped cylindrical Fermi surface the dispersion of the miniband along $k_z$ is weak. The miniband is then filled within a narrow field interval above $B^*$, and the maximum of the zero-bias conductance is higher and narrower than for the spherical Fermi surface considered here. The limit of a two-dimensional film corresponds to a level crossing at $B^*$ without Fermi points. The third effect is spin-orbit coupling, which makes $g_{\rm eff}$ a band-dependent quantity and removes the conservation of the spin projection. The linear splitting of the levels remains, and its slope measures $g_{\rm eff}$ for the pocket that forms the core states. The fourth effect is the vortex lattice. At $B=3.5$\,T the distance between vortices is $26$\,nm, a few times $\xi$, and tunneling between the cores adds to the width of the levels. Finally, the instability of the 1D Fermi liquid considered in Ref.~\cite{MakhlinVolovik95} would open a gap at the Fermi points below a transition temperature that is exponentially small for weak coupling. We return to this point in Sec.~\ref{sec:int}.

The field dependence of vortex spectra in Fe(Te,Se) was measured between $1$ and $6$\,T with a resolution of $20\,\mu$eV in Ref.~\cite{Machida19}. The spectra of the same vortex at $1$ and $3$\,T were found to be different, and the Zeeman effect was named as an open theoretical problem. The expected shift of a level between these fields is $0.12$\,meV for $g_{\rm eff}=2$, several times the resolution. A direct test of the present theory is the field dependence of the lowest level in a vortex without a zero-bias peak.

\section{Interaction in the filled miniband}
\label{sec:int}

\subsection{Projected interaction}

Above the threshold the low-energy degrees of freedom are the spin-up quasiparticles of the lowest miniband with momenta near $\pm\kf$. We denote their annihilation operators by $\gamma_k$. The other minibands are separated from the Fermi level by energies of the order of $E_0$. The electron density operator projected onto the lowest miniband is \cite{VishwanathSenthil01}
\begin{equation}
\rho(\mathbf{r})\to\frac1L\sum_{k,k'}e^{i(k'-k)z}\,Q_{kk'}(r)\,\gamma_k^\dagger\gamma_{k'},
\label{eq:rho}
\end{equation}
where $Q_{kk'}=f_kf_{k'}-g_kg_{k'}$.
The quasiparticle couples to the density through its charge profile $Q_{kk}=f_k^2-g_k^2$. For a density-density interaction $V(\mathbf{r}-\mathbf{r}')$ the projected Hamiltonian is
\begin{equation}
H_{\rm int}=\frac{1}{2L}\sum_{k_1,k_2,q}V_q(k_1,k_2)\,\gamma^\dagger_{k_1+q}\gamma^\dagger_{k_2-q}\gamma_{k_2}\gamma_{k_1},
\label{eq:Hint}
\end{equation}
with the matrix element
\begin{align}
V_q(k_1,k_2)=\!\int\! d^2\rho\,d^2\rho'\;&V_q(\bm\rho-\bm\rho')\nonumber\\
&\times Q_{k_1,k_1+q}(\rho)\,Q_{k_2,k_2-q}(\rho').
\label{eq:Vq}
\end{align}
Here $V_q(\bm\rho)$ is the Fourier transform of the interaction along $z$. The matrix element depends on the momentum transfer $q$ and also on the momenta of the two particles, because the radial functions depend on $k_z$ through $k_\perp$. The same structure appears at a quantum Hall edge, where the matrix element depends on the guiding-center coordinates of the two electrons \cite{Karzig12}.

Inside the superconductor the interaction is screened. We use the Thomas--Fermi form with the screening wave vector $\qTF$, $\qTF^2=e^2N_0/\epsilon_0\epsilon_r$, where $\epsilon_r$ is the background dielectric constant. Its range $1/\qTF$ is shorter than the period $\pi/k_\perp$ of the oscillations of $Q$ when $\qTF>2\kF$. The kernel can then be replaced by a local one with the same integral,
\begin{equation}
V_q(k_1,k_2)\simeq\frac{1}{N_0}\,\frac{S(k_1,k_2)}{1+q^2/\qTF^2},
\label{eq:local}
\end{equation}
with the overlap $S(k_1,k_2)=\int d^2\rho\,Q_{k_1k_1}(\rho)\,Q_{k_2k_2}(\rho)$. Here we neglected the difference between $Q_{k,k+q}$ and $Q_{kk}$ in the overlap. We evaluate $S$ with the numerical BdG functions. The diagonal value $S_2(k)\equiv S(k,k)$ increases from $0.175/\xi^2$ at $k=0$ to $0.23/\xi^2$ at $k=0.8\,\kF$. The net charge of the quasiparticle is small, $\int d^2\rho\,Q_{kk}=0.015$ at $k=0$, so the coupling comes from the oscillating part of the charge profile.

\subsection{Luttinger parameter}

For spin-polarized fermions the contact part of the interaction drops out. The coupling that enters the low-energy theory is the difference between the forward and the backward matrix elements at the Fermi points,
\begin{equation}
U=V_0(\kf,-\kf)-V_{2\kf}(\kf,-\kf)=\frac{S_2(\kf)}{N_0}\,\frac{4\kf^2}{\qTF^2+4\kf^2}.
\label{eq:U}
\end{equation}
The two Fermi points have the same $k_\perp$ and the same radial functions, so the two terms differ only through the screening factor. The dimensionless coupling and the Luttinger parameter are
\begin{equation}
g_{1D}=\frac{U}{2\pi\hbar\vf},\qquad K=\frac{1}{\sqrt{1+2g_{1D}}} .
\label{eq:K}
\end{equation}
It is convenient to introduce the parameter
\begin{equation}
\lambda=\frac{S_2(\kf)}{\pi\hbar\vf N_0}=\frac{\pi\,S_2\xi^2}{\kF\xi}\,\frac{\Delta}{\hbar\vf\kF},
\label{eq:lambda}
\end{equation}
in terms of which $g_{1D}=(\lambda/2)\,4\kf^2/(\qTF^2+4\kf^2)$.

Figure~\ref{fig:params}(c) shows $\lambda$ and $g_{1D}$ as functions of the field. The parameter $\lambda$ is large near the threshold, where the velocity is small, and decreases to $0.55$ at $1.5\,B^*$ and $0.23$ at $2\,B^*$. The coupling $g_{1D}$ contains the additional factor $4\kf^2/\qTF^2$, which vanishes at the threshold. As a result $g_{1D}$ has a maximum at $B\approx1.15\,B^*$. The maximal value is $0.14$ for $\qTF=2\kF$, $0.07$ for $\qTF=3\kF$, and $0.04$ for $\qTF=4\kF$. The Luttinger parameter stays between $0.88$ and $1$. In a multiband material the screening involves the total density of states, which exceeds $N_0$ of a single band and reduces these values further.

The tunneling exponent for the end of a Luttinger liquid is $1/K-1\simeq g_{1D}$ \cite{KaneFisher92}, and the bulk exponent is of the order of $g_{1D}^2/2$. With $g_{1D}\sim0.1$ or smaller the power-law suppression of the tunneling density of states at the Fermi level \cite{VishwanathSenthil01} is a weak correction to the single-particle result \eqref{eq:zb} within the available range of energies between $w$ and $\ef$. The filled miniband is thus a weakly interacting 1D Fermi gas. The spectra of Sec.~\ref{sec:stm}, which follow from the single-particle theory, are its main observable.

The same conclusion applies to the pairing instability of Ref.~\cite{MakhlinVolovik95}. For weak coupling the corresponding transition temperature is exponentially small in the inverse coupling constant of the triplet channel, and the gapless description holds at accessible temperatures.

\section{Relaxation by three-particle collisions}
\label{sec:kin}

\subsection{Background}

Two-particle collisions in a single 1D band conserve the momenta of the colliding particles up to a permutation and do not relax the distribution. Relaxation in 1D systems is governed by three-particle collisions \cite{Lunde07,Imambekov12,LevchenkoMicklitz21}. In quantum wires these processes determine the decay rate of quasiparticles \cite{Khodas07,Karzig10,MatveevFurusaki13}, the thermalization of the electron distribution \cite{MicklitzLevchenko11}, and the equilibration between right- and left-moving particles, which proceeds through the diffusion of holes near the bottom of the band \cite{Micklitz10,Levchenko10,MatveevAndreev12}. Relaxation of this type was observed in momentum-resolved tunneling between quantum wires \cite{Barak10}. For the chiral edge channels of the integer quantum Hall effect, energy relaxation was measured in Refs.~\cite{Altimiras10,leSueur10}, and the three-particle mechanism was analyzed in Ref.~\cite{Karzig12}. There the dependence of the interaction matrix element on the positions of the guiding centers gives a finite amplitude.

For spinless fermions with a contact interaction the three-particle amplitude vanishes, because the contact interaction has no effect on identical fermions. A finite amplitude requires a dependence of the matrix element on momenta. Matveev and Furusaki showed that for spinless fermions near the Fermi points the amplitude has a universal form fixed by the statistics, and that the decay rate of a quasiparticle at zero temperature scales as the eighth power of its energy \cite{MatveevFurusaki13}. The filled vortex miniband realizes this case. Its specific feature is the matrix element \eqref{eq:Vq}, which depends on the momenta of the particles through the radial functions.

\subsection{Three-particle amplitude}

\begin{figure*}[t]
\includegraphics[width=0.9\textwidth]{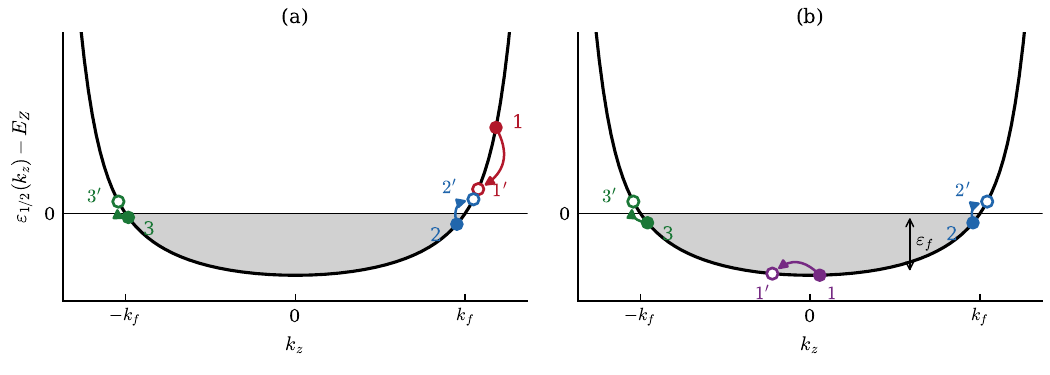}
\caption{Three-particle collisions in the filled miniband, drawn on the numerical dispersion at $B=1.6\,B^*$. Filled and open circles are the initial and final states. (a) Decay of a quasiparticle (1) above the Fermi level. A second particle (2) is excited at the same Fermi point, and a third particle (3) at the opposite Fermi point takes up the small momentum mismatch. (b) Scattering of a hole near the bottom of the band by particles at the two Fermi points. This process controls the equilibration between right- and left-moving particles.}
\label{fig:kin}
\end{figure*}

We consider the decay of a spin-up quasiparticle with momentum $k_1>\kf$ and excitation energy $\varepsilon=\hbar\vf(k_1-\kf)\ll\ef$ at $T=0$. Momentum and energy conservation allow the process shown in Fig.~\ref{fig:kin}(a). The hot particle transfers the momentum $|q_1|<k_1-\kf$ to a particle near the right Fermi point, and a particle near the left Fermi point takes up the momentum $|q_3|\simeq|q_1|(k_1-\kf-|q_1|)/2\kf\ll|q_1|$.

The amplitude $A$ of the transition $(k_1,k_2,k_3)\to(k_1',k_2',k_3')$ is given by second-order perturbation theory in $H_{\rm int}$, antisymmetrized over the final states (Appendix~\ref{app:three}). We expand the matrix element \eqref{eq:local} near the Fermi points. The overlap $S(k_1,k_2)$ is an even function of each momentum. Its expansion contains terms that depend on $k_1^2$ and $k_2^2$ separately and a term proportional to $(k_1^2-k_2^2)^2$,
\begin{equation}
V_q(k_1,k_2)\simeq V_0\,\frac{1-\beta\,(k_1^2-k_2^2)^2}{1+q^2/\qTF^2},
\label{eq:Vmodel}
\end{equation}
where $V_0=S_2(\kf)/N_0$ and the separable terms are omitted. The coefficient $\beta$ is determined by the mixed second derivative of $S$ with respect to $k_1^2$ and $k_2^2$. We evaluated the amplitude numerically with high precision for the parabolic dispersion $\hbar^2k^2/2m^*$ and for several forms of the matrix element. The results are as follows. The amplitude vanishes for a constant matrix element, for a matrix element that depends on $k_1^2$ and $k_2^2$ separately, and for a matrix element of the form $V_0(1-cq^2)$. The contributions of first order in $\beta$ and in the coefficient of the $q^4$ term also vanish. The leading contribution is of second order in the momentum-dependent part of the matrix element and has the form
\begin{equation}
A=\frac{\Lambda}{L^2}\,(k_1-k_2)(k_1'-k_2'),
\label{eq:A}
\end{equation}
where $k_1,k_2$ and $k_1',k_2'$ are the initial and final momenta of the two particles at the same Fermi point. This is the structure derived in Ref.~\cite{MatveevFurusaki13}. For the constant we find
\begin{equation}
\Lambda=\frac{32\,m^*V_0^2\kf^4}{\hbar^2}\Big(\beta^2-\frac{2\beta}{\qTF^4}+\frac{9}{2\,\qTF^8}\Big).
\label{eq:Lambda}
\end{equation}
The numerical coefficients $-2$ and $9/2$ are accurate to three digits. The amplitude with the first term alone is known in closed form for an arbitrary momentum of the hot particle (Appendix~\ref{app:three}).

The numerical BdG functions give $\beta\kF^4=2.4$ at $\kf\to0$ and $3.2$ at $\kf=0.6\,\kF$. For $\qTF\ge2\kF$ one has $\beta\,\qTF^4\ge38$, and the screening terms in Eq.~\eqref{eq:Lambda} change $\Lambda$ by less than $6\%$. The amplitude is therefore dominated by the dependence of the radial functions on $k_z$. We characterize it by the dimensionless parameter
\begin{equation}
\eta=\beta\,\kf^4 ,
\label{eq:eta}
\end{equation}
which is shown in Fig.~\ref{fig:params}(c). It is small near the threshold and reaches $0.8$ at $1.5\,B^*$.

\subsection{Relaxation rates}

The golden rule with the amplitude \eqref{eq:A} gives the decay rate of the hot quasiparticle (Appendix~\ref{app:three})
\begin{equation}
\frac1\tau=\frac{3\,\Lambda^2\,\varepsilon^8}{5120\,\pi^3\hbar^9\,m^*\vf^{10}},\qquad T\ll\frac{\varepsilon^2}{\ef}.
\label{eq:rateMF}
\end{equation}
This expression coincides with the result of Ref.~\cite{MatveevFurusaki13}, and Eq.~\eqref{eq:Lambda} supplies the constant $\Lambda$ for the vortex. For the parabolic band, $\vf=\hbar\kf/m^*$ and $\ef=\hbar^2\kf^2/2m^*$, and with the first term of Eq.~\eqref{eq:Lambda} the rate becomes
\begin{equation}
\frac1\tau=\frac{3\pi}{640}\,\eta^4\lambda^4\,\frac{\varepsilon^8}{\hbar\,\ef^7}.
\label{eq:rate}
\end{equation}
The rate is of fourth order in the interaction, because the amplitude is of second order in the non-contact part of the matrix element. At temperatures $\varepsilon^2/\ef\ll T\ll\varepsilon$ the phase space of the particle at the opposite Fermi point is set by the temperature, and
\begin{equation}
\frac1\tau=\frac{11\pi}{180}\,\eta^4\lambda^4\,\frac{T\,\varepsilon^6}{\hbar\,\ef^6}.
\label{eq:rateT}
\end{equation}
Here and below the temperature is measured in energy units. For thermal quasiparticles, $\varepsilon\sim T$, the rate scales as $T^7$.

Equilibration between the right- and left-moving particles requires backscattering. It proceeds through the diffusion of a hole in momentum space near the bottom of the band, as shown in Fig.~\ref{fig:kin}(b) \cite{Micklitz10,MatveevAndreev12}. The amplitude for a hole at the bottom is constant at small momentum transfer and equals $A_h=\Lambda\kf^2/4L^2$. The momentum diffusion coefficient of the hole and the corresponding relaxation rate are
\begin{subequations}\label{eq:hole}
\begin{align}
\mathcal{B}&=\frac{256\pi^5}{15}\,\eta^4\lambda^4\,\frac{m^{*4}T^5}{\hbar^7\kf^6},\\
\frac{1}{\tau_h}&=\frac{\mathcal{B}}{m^*T}=\frac{32\pi^5}{15}\,\eta^4\lambda^4\,\frac{T^4}{\hbar\,\ef^3}.
\end{align}
\end{subequations}
The equilibration rate contains in addition the probability to find a hole at the bottom of the band, which is proportional to $e^{-\ef/T}$. In semiconductor quantum wires the ratio $\ef/T$ is large and equilibration is exponentially slow. In the vortex $\ef=E_0(B/B^*-1)$ is tunable by the field and is comparable to the temperature near the threshold, so the activation factor is not small.

\subsection{Estimates and other relaxation channels}

Equation~\eqref{eq:rate} assumes a parabolic band, which holds for $\ef\ll E_0$. At higher fields the band is not parabolic. For an estimate we use Eq.~\eqref{eq:rateMF} with the numerical values of $\vf$ and of the band curvature at the Fermi point. The result is shown in Fig.~\ref{fig:params}(d). For $\varepsilon=\ef$ the rate has a maximum $\hbar/\tau\approx10^{-6}E_0$ at $B\approx1.3\,B^*$. With $E_0=0.2$\,meV this corresponds to $\tau$ of the order of a microsecond. The rate decreases as $\varepsilon^8$ at lower energies. The use of the single-band density of states in $V_0$ overestimates $\lambda$ in a multiband material, so these values are upper bounds for the rates. Inelastic electron collisions thus give a negligible contribution to the width of the levels, and the 1D gas in the core is collisionless on the time scales of other processes.

Emission of acoustic phonons by a quasiparticle of the miniband requires its velocity to exceed the sound velocity. The velocity scale is $E_0/\hbar\kF$, which equals $0.3$\,km/s for $E_0=0.2$\,meV and $\kF=0.1$\,\AA$^{-1}$. According to Fig.~\ref{fig:params}(b), the Fermi velocity stays below a typical sound velocity of a few km/s up to $B\approx2\,B^*$. In this range the emission of a single phonon by a quasiparticle near the Fermi points is forbidden kinematically. The coupling to long-wavelength phonons is further reduced by the neutrality of the bound state. Impurities scatter quasiparticles elastically between the two Fermi points and determine the level width observed in spectroscopy \cite{Room26}. Quasiparticles injected by the STM tip propagate along the vortex line away from the surface with the velocity $\vf$, so the slow relaxation does not lead to their accumulation under the tip.

\subsection{Relation to vortex dynamics}

The relaxation of core quasiparticles is an input of the theory of flux flow. In the quasiclassical theory developed in Refs. \cite{KopninKravtsov76,Kopnin01} the forces on a moving vortex depend on the parameter $\omega_0\tau$, where $\tau$ is the relaxation time of the core quasiparticles. This theory treats the spectrum of the core as continuous. Authors of Ref. \cite{FeigelmanSkvortsov97} considered a layered superconductor with moderate disorder at low temperatures, where the inelastic width of the core levels is smaller than their spacing. Dissipation then comes from Landau--Zener transitions between the levels of a moving vortex, followed by the emission of phonons. In their analysis the inelastic width is attributed mainly to the electron-phonon interaction.

The present work concerns the regime $T\ll E_0$ in a clean three-dimensional superconductor, where the levels are minibands and one of them is partially filled. The rates \eqref{eq:rateMF}--\eqref{eq:hole} describe the redistribution of energy and momentum among the quasiparticles within this miniband. These collisions conserve the total energy, the total momentum along the vortex, and the angular momentum of the quasiparticles. They establish a local equilibrium of the 1D gas but do not transfer energy or angular momentum to the lattice, and they do not contribute to the friction force on the vortex. Our estimates show that electron collisions within the core are slow in the quantum limit, which supports the neglect of this channel in the theory of flux flow at low temperatures. A theory of flux flow above the threshold field has to account for a new element, the gapless excitations of the filled miniband. This problem is left for future work.

\section{Summary}
\label{sec:sum}

We studied the field-induced Lifshitz transition in the core of a superconducting vortex, at which the lowest CdGM miniband of one spin projection crosses the Fermi level. The idea of this transition goes back to Refs.~\cite{MakhlinVolovik95,VishwanathSenthil01}. Our results are the following.

The tunneling conductance was obtained from the numerical solution of the BdG equations in the quantum limit. The peak at the vortex center splits linearly with the field. The zero-bias conductance at the center appears at the threshold field $B^*=2E_0/g_{\rm eff}\mu_B$, reaches a maximum of the order of $6\,N_0(E_0/w)^{1/2}$, and decays as $(B-B^*)^{-1/2}$. The zero-bias signal is spin-up polarized at the center and spin-down polarized in the surrounding ring. The miniband with $\mu=3/2$ gives a second onset in the ring. The transition is resolved when the total broadening is a fraction of the minigap $E_0$.

The interaction between the quasiparticles of the filled miniband was evaluated with the BdG wave functions. The dimensionless coupling is of the order of $0.1$ or smaller for realistic screening, and the Luttinger parameter is close to unity. Quasiparticles relax by three-particle collisions. The amplitude is determined by the dependence of the transverse wave functions on the momentum along the vortex, and the decay rate \eqref{eq:rateMF} scales as $\varepsilon^8$. The corresponding relaxation times are of the order of microseconds or longer.

FeSe is a suitable material for the observation of the transition. The measured minigap $E_0\approx0.2$\,meV gives $B^*\approx3.5$\,T for $g_{\rm eff}=2$, which is within the range of existing dilution-refrigerator STM instruments and below $H_{c2}$.

Two extensions are of interest. The first is a model with several bands, an anisotropic gap, and a self-consistent order parameter. The second is the role of spin-orbit coupling. Without conservation of the spin projection the two Fermi points of the filled miniband can be coupled by pairing. The resulting state of the vortex line and its possible end states are an open question.

\begin{acknowledgments}
This work was supported by the U.S. Department of Energy (DOE), Office of Science, Basic Energy Sciences (BES) under Award No. DE-SC0020313. The author acknowledges H. I. Romnes Faculty Fellowship provided by the University of Wisconsin-Madison Office of the Vice Chancellor for Research and Graduate Education with funding from the Wisconsin Alumni Research Foundation. 

The author used the large language model Claude (Anthropic, model Claude Fable 5.2) in this work. Under the author's direction, the model wrote and ran the code for the numerical solution of the Bogoliubov--de Gennes equations and the tunneling spectra. It produced the figures from these calculations and assisted with the literature search. The kinetic analysis follows the author's earlier work on one-dimensional systems. The author conceived and directed the project, verified the results, and takes full responsibility for the content of the paper.
\end{acknowledgments}

\appendix

\section{Numerical solution of the BdG equations}
\label{app:num}

Equations~\eqref{eq:radial} are written in units of $\Delta$ and $\xi$. The kinetic prefactor is $\hbar^2/2m\xi^2\Delta=1/2\kF\xi$, and the transverse Fermi energy is $E_\perp/\Delta=(\kF\xi/2)(1-k_z^2/\kF^2)$. The radial interval $0<r\le12\,\xi$ is discretized with $1200$ points. The operator $r^{-1}\partial_r(r\partial_r)$ is represented in the conservative form on the half-integer grid and made symmetric by the substitution $f\to\sqrt{r}f$. For the component with zero orbital momentum the boundary condition at $r=0$ is $f'=0$. The eigenvalues closest to zero are found by the shift-invert Lanczos method. We keep all states with $|E|<1.05\,\Delta$ for $\mu=\pm1/2,\dots,\pm13/2$ at $124$ values of $k_z$ between $0$ and $0.999\,\kF$. Between neighboring values of $k_z$ the energies and the densities $f^2$ and $g^2$ of states with the same $\mu$ and the same index are interpolated linearly, so that the step in energy does not exceed $0.04\,E_0$. The integral over $k_z$ in Eq.~\eqref{eq:ldos} is then evaluated by the midpoint rule. For $\kF\xi=12$ we used $1800$ radial points. The symmetry $E\to-E$, $\mu\to-\mu$ holds to the numerical precision.

For $\kF\xi=4$, $6$, and $12$ the minigap is $E_0/\Delta=0.114$, $0.073$, and $0.036$, and the ratios $\varepsilon_{3/2}(0)/E_0$ and $\varepsilon_{5/2}(0)/E_0$ are $(2.80,4.23)$, $(2.92,4.68)$, and $(2.95,4.88)$. The product $2E_FE_0/\Delta^2$ equals $0.46$, $0.44$, and $0.43$, close to the quasiclassical value $I(1)/2=0.43$.

The overlap $S(k_1,k_2)$ of Eq.~\eqref{eq:local} is computed from the functions of the lowest miniband on the full radial grid. The coefficient $\beta$ is obtained from the combination $S(k_1,k_2)-[S(k_1,k_1)+S(k_2,k_2)]/2=-\beta\,S_2\,(k_1^2-k_2^2)^2$, which removes the separable terms. The result changes by less than $5\%$ when the distance between $k_1$ and $k_2$ is varied between $0.05\,\kF$ and $0.17\,\kF$.

\section{Three-particle amplitude and rates}
\label{app:three}

\subsection{Amplitude}

We define the two-particle vertex for the transition $(a,b)\to(a',b')$ as $v=V_{a'-a}\big(\tfrac{a+a'}{2},\tfrac{b+b'}{2}\big)$. To second order in $H_{\rm int}$ the amplitude for three distinguishable particles is a sum of six terms. In each term two particles $i$ and $j$ collide first, particle $i$ goes to its final state, and particle $j$ goes to an intermediate state $k_j''=k_i+k_j-k_i'$ and then collides with the third particle $s$,
\begin{equation}
A_{\rm dir}=\frac{1}{L^2}\sum_{(i,j)}\frac{v(k_i,k_j\to k_i',k_j'')\,v(k_j'',k_s\to k_j',k_s')}{\varepsilon_{k_i}+\varepsilon_{k_j}-\varepsilon_{k_i'}-\varepsilon_{k_j''}} .
\end{equation}
For identical fermions $A$ is the sum of $A_{\rm dir}$ over the six permutations of the final momenta with the sign of the permutation. The transition rate is $(2\pi/\hbar)|A|^2\delta(E_i-E_f)$ with momentum conservation \cite{Lunde07,LevchenkoMicklitz21}.

Each of the 36 terms diverges at small momentum transfer, and the result comes from their cancellation. We evaluated the sum analytically on the energy shell for $\varepsilon_k=\hbar^2k^2/2m^*$. For the matrix element $V_0[1-\beta(k_a^2-k_b^2)^2]$ with $k_2=\kf$ and $k_3=-\kf$ the amplitude at small $q_1$ is
\begin{equation}
A=\frac{8\,\beta^2V_0^2m^*}{\hbar^2L^2}\,\kf^2\,(k_1^2-\kf^2)^2\Big(1-\frac{2|q_1|}{k_1-\kf}\Big).
\label{eq:Afull}
\end{equation}
The prefactor was verified nmerically for several values of $\kf$, $k_1$, $m^*$, and $\beta$. For $k_1\to\kf$ Eq.~\eqref{eq:Afull} reduces to Eq.~\eqref{eq:A} with the first term of Eq.~\eqref{eq:Lambda}, because $k_1'-k_2'\simeq k_1-\kf-2|q_1|$. For $k_1\to0$ it gives the amplitude $8\beta^2V_0^2m^*\kf^6/\hbar^2L^2$ for a hole at the bottom of the band. The coefficients of the screening terms in Eq.~\eqref{eq:Lambda} were obtained from the limit $k_1\to\kf$, $q_1\to0$ for the matrix elements $V_0[1-\beta(k_a^2-k_b^2)^2]/(1+q^2/\qTF^2)$ at small $\beta\kf^4$ and $\kf/\qTF$.

\subsection{Decay rate}

The decay rate of the particle with momentum $k_1$ is
\begin{align}
\frac1\tau=\frac{\pi}{\hbar}\!\sum_{k_2,k_3,q_1,q_2}\!&|A|^2\,\delta(E_i-E_f)\nonumber\\
&\times n_{k_2}(1-n_{k_2'})\,n_{k_3}(1-n_{k_3'}),
\end{align}
where $n_k$ is the occupation number, $k_3'=k_3-q_1-q_2$, and the prefactor $\pi/\hbar$ includes the factor $1/2$ that accounts for the identity of the two final particles at the right Fermi point. We set $|q_1|=q$ and $k_2=\kf-p$. For momenta near the Fermi points $E_i-E_f=-\hbar[(v_1+\vf)q_1+2\vf q_2]$, so the integration over $q_2$ gives $1/2\hbar\vf$. At $T=0$ the occupation numbers restrict $0<p<q<k_1-\kf$, and the sum over $k_3$ gives the factor $|q_3|=q(k_1-\kf-q+p)/2\kf$. With $A=(\Lambda/L^2)(k_1-\kf+p)(k_1-\kf-2q+p)$ we obtain
\begin{equation}
\frac1\tau=\frac{\Lambda^2(k_1-\kf)^8}{64\pi^3\hbar^2\vf\kf}\,J,
\end{equation}
\begin{equation}
J=\int_0^1\!du\!\int_0^u\!dw\,(1+w)^2(1-2u+w)^2\,u\,(1-u+w)=\frac{3}{80}.
\end{equation}
Substitution of $k_1-\kf=\varepsilon/\hbar\vf$ and $\kf=m^*\vf/\hbar$ gives Eq.~\eqref{eq:rateMF}. At $\varepsilon^2/\ef\ll T\ll\varepsilon$ the factor $|q_3|$ is replaced by $T/\hbar\vf$, the integral becomes $\int_0^1du\int_0^udw\,(1+w)^2(1-2u+w)^2=11/90$, and Eq.~\eqref{eq:rateT} follows.

For a hole near the bottom of the band, energy conservation gives $q_2\simeq q_3\simeq-q_1/2$. The rate of the transition with the momentum change $q_1$ contains the factor $(q_1/2)^2\,b(1+b)$, where $b=[\exp(\hbar\vf|q_1|/2T)-1]^{-1}$. The diffusion coefficient $\mathcal{B}=\sum_{q_1}(\hbar q_1)^2W(q_1)$ then equals $(4\pi/15)\,|A_hL^2|^2\,T^5/\hbar^5\vf^6$, where $A_h$ is the hole amplitude given above. This leads to Eq.~\eqref{eq:hole}.

\bibliography{refs}

\end{document}